\documentstyle{article}
\begin{document}
\begin{center}
{\Large  Symmetry as a source of hidden coherent structures in quantum
physics: general outlook and examples} \\ [0.1cm]
Valery P. Karassiov\\ [0.1cm]
P.N. Lebedev Physical Institute, Leninsky Pr. 53, Moscow, 117924 Russia
\end{center}
\begin{abstract}
A general algebraic approach, incorporating both invariance groups and
dynamic symmetry algebras, is developed to reveal hidden coherent structures
(closed complexes and configurations) in quantum many-body physics models
due to symmetries of their Hamiltonians $H$.
Its general ideas are manifested on some recent new examples:
1) $G$-invariant bi-photons and a related $SU(2)$-invariant treatment of
unpolarized light; 2) quasi-spin clusters in nonlinear models of quantum
optics; 3) construction of composite particles and (para)fields from
$G$-invariant clusters due to internal symmetries.

\end{abstract}

\section{Introduction. General remarks }
 
The symmetry methods are widely used in quantum physics from the time of its
origin and up to now because they yield powerful epistemological and
computational tools for examining many physical problems (see, e.g., [1-14]
and literature cited therein). In particular, invariance principles provide
formulations of dynamic laws and classifications of quantum states which are
most adequate to reveal different physical phenomena \cite{w1,wi} whereas
the formalism of groups and Lie algebras, especially, generalized coherent
states and related techniques, yield simple and elegant solutions of
spectral and evolution problems \cite{per,k91}. From the spectroscopic point
of view one distinguishes two (exploiting, as a rule, independently) types of
physical symmetries depending on the behaviour of Hamiltonians $H$ under
study with respect to symmetry transformations \cite{bar}. One of them,
associated with invariance groups $G_i(H)\, ([G_i,H]_-=0)$ of
Hamiltonians, describes (non-accidental) degeneracies of energy spectra
within fixed irreducible representations (IRs) of $G_i(H)$ while another one,
connected with so-called dynamic symmetry (or spectrum generating) algebras
$g^{D}\, ([g^{D},H]_-\subseteq g^{D}\neq 0)$ \cite{bar}, enables to determine
such spectra within fixed IRs of $g^{D}$ and to give spectral decompositions
of Hilbert spaces $L(H)$ of quantum systems in $g^{D}$-invariant subspaces
$L(\lambda)$ (with $\lambda$ being labels of $g^{D}$ IRs $D^{\lambda}$)
which describe certain (macroscopic) coherent structures (CS), i.e., stable
sets of states (shells, (super)multiplets, configurations, phases,
etc. [3-12]) evolving in time independently under actions of $H$.

Applications of these methods are especially fruitful in examining
many-body problems whose Hamiltonians $H$ and quantum state spaces $L(H)$ are
given in terms of boson-fermion operators: $H=H(a_i, a_i^+, b_j, b_j^+), \,
L(H)\subseteq L_F(n;m)\equiv Span\{\prod_{i=1}^{n}(a^+_i)^{\eta_i}
\prod_{j=1}^{m}(b^+_j)^{\mu_j}|0\rangle\} \;([a_i, a_j^+]_-=\delta_{ij}=
[b_i,b_j^+]_+)$. Indeed, various (originated from the works \cite{jhp})
boson-fermion mappings $f$:
\begin{equation}(a_i, a_i^+, b_j, b_j^+)\;\stackrel{f}\longmapsto\;
F_{\alpha},\qquad [F_{\alpha},
F_{\beta}]_{\mp}\equiv F_{\alpha} F_{\beta} \mp F_{\beta} F_{\alpha} =
\sum_{\gamma =1}^d c_{\alpha \beta}^{\gamma}F_{\gamma} \equiv
\psi_{\alpha \beta}(\{F_{\alpha}\})
\label{eq:bfm}
\end{equation}
enable us to introduce generators $F_{\beta}$ of $d (<\infty)$-dimensional
Lie algebras (or superalgebras \cite{om}) as (super)symmetry operators of
both types and collective dynamic variables of problems under study in whose
terms one gets reformulations of $H, L(H)$ facilitating solutions of many,
mainly, spectrocopic many-body tasks [3-12]. On the other hand, within
many-body models, due to composite structures of their "elementary"
coupled micro-objects (quasi-particles, clusters, etc.), one can reveal in
a natural manner deep (although hidden) interrelations between both symmetry
types above, and, therefore, a study of one of them automatically yields
an information about other one\cite{ksh}. A consequent realization of this
standpoint, being complemented by an "invariant confinement principle" (for
constituents), leads to an unified "invariant-dynamic" approach (IDA) to
reveal new cooperative effects and phenomena in many-body physics on both
micro- and macro-levels \cite{k88}; from the methodological point of view
it may be considered as a specification of the general nature-philosophical
principle (used at the intuitional level already by I. Kepler and explicitly
realized by H. Weyl and E. Wigner within modern physics \cite{w1,wi}):
\,{\it symmetry generates (induces) a formation of CS (coherent
configurations) in sets of interacting objects}.

Note that single (mostly, formal) aspects of IDA were implemented in quantum
physics long ago, beginning from using binary $SU(2)$-invariants to describe
bi-polar molecular valent bonds in 1931 \cite{rtw} and from group-theoretical
studies of complex atom spectra by G. Racah in 1942-49 \cite{rg}.
Specifically, the latters were fruitfully  developed later in nuclear,
atomic and molecular physics \cite{ed,li,bl1,nad} and have led to exact
definitions of two basic concepts of IDA \cite{k91,k93a}: dynamic symmetry
algebras (see \cite{li,bar} and references therein) and complementarity of
groups \cite{mq}. The most complete (although implicit) implementations of
IDA were given in superfluidity/superconductivity theories by introducing
$U(1)$-invariant Bogolubov's/Cooper's pairs and the $SU(1,1)/SU(2)$
canonical transformations \cite{bts} and in  particle physics (interrelations
between "colour" and "flavour" $SU(n)$ symmetries) \cite{ok}. However, up to
recently an explicit mathematical formulation of IDA, summing up such
implementations, was absent that prevented its systematic applications. The
aim of the paper is to give (destined
for physicists) mathematical grounds of IDA \cite{k91,k97} and to manifest
its efficiency and physical meaning on some recently examined examples in
quantum optics [20,24-27] and in the theory of composite particles and
fields with internal symmetries \cite{k91,k97}.

The paper is dedicated to the memory of Academician N.N. Bogolubov, whose
ideas and works promoted to forming IDA, and of Professor Ya. A.
Smorodinsky, discussions with whom stimulated the developments
presented below.

\section{Mathematical grounds:  $G$-invariant Jordan mappings and
Weyl-Howe dual pairs in many-body physics}

The mathematical formulation of IDA is based on a synthesis of vector 
invariant theory \cite{w1,how} and extensions \cite{k91,k93a} of the concept
of complementary groups and of the Jordan mapping \cite{jhp}.

As is known, the original Jordan mapping, given by Eq. (\ref{eq:bfm}) with 
quadratic functions $f$, introduces collective dynamic variables 
$F_{\alpha}(t)$ related to generators $F_{\alpha}$ of certain Lie
(super) algebras $g_0^D$ of dynamic symmetry and reduces quadratic (in field
operators) Hamiltonians $H_0(a_i, a_i^+, b_j, b_j^+)$, describing free and
linear (in the Heisenberg picture) dynamics, to the form
\begin{equation}
H_0=\sum_{\alpha} \lambda_{\alpha} F_{\alpha}+C,\quad [F_{\alpha},C]_- =0,
\quad  Span\{F_{\alpha}\}= g_0^D
\label{eq:H0}
\end{equation}
where $\lambda_{\alpha}$ are $c$-number coefficients; herewith algebras
$g_0^D$ for particular $H$ are subalgebras of certain "maximal" (in a sense)
finite-dimensional Lie superalgebras $g_0^{DM}$ which act on $L_F(n;m)$
irreducibly and are semi-direct products of the superalgebras $osp(2n|2m)$
(with the even part $sp(2n,R)\otimes o(2m)$) and the Weyl-Heisenberg
superalgebras $w(n,m)= Span(a_i, a_i^+, b_j, b_j^+)$ \cite{bar,om,nad,k91}.

Suppose now that $H_0$ have (both continious and discrete) invariance groups
$G_i(H_0)=G_i^0$ and field operators form sets of vectors $a_i^+=(a_{ki}^+),
b_j^+ =(b_{kj}^+)$ which are transformed with respect to some (e.g.,
fundamental) IRs of the groups $G_i^0$. Then $F_{\alpha}\in g_0^D$ are
quadratic vector $G_i^0$-invariants and, besides, there is the
characteristic equation: $[g_0^D,\; G_i ^0)]_-=0 \,
\Longleftrightarrow \, G_i^0 \,g_0^D \,G^{0-1}_i= g_0^D$
entailing functional connections between $G_i$- and $g_0^D$-invariant
(Casimir and class \cite{ed}) operators $C_k(G_i)$ and $C_k(g_0^D)$ and
specifications of their eigenvalues on spaces $L(H)$ by common sets
$[l_i]\equiv [l_0,l_1,\dots]$ of invariant quantum numbers $l_i$ which
determine IRs of both $G_{i}$ and $g_0^D$ and label their common extremal
(usually, lowest) vectors $|[l_i]\rangle$ \cite{k97}. All that, in turn,
yields spectral decompositions
\begin{equation}
L(H) =\sum_{[l_i]}\;\sigma([l_i])\; L([l_i]),\quad L([l_i])=
 Span\{D^{[l_i]}(G^0_i)\otimes D^{[l_i]}(g_0^D)|[l_i]\rangle\}
\label{eq:SD}
\end{equation}
of spaces $L(H)$ in direct sums of the subspaces $L([l_i])$ which are
invariant with respect to joint actions $D(g_0^D)\otimes D(G_i^0)$ of
algebras $g_0^D$ and groups $G_i^0$ being carrier-spaces of so-called
isotypic components (factor-representations) \cite{bar} of both these
algebraic structures, i.e. $L([l_i])$ contains carrier-spaces of equivalent
IRs $D^{[l_i]}(G_i^0)\, (D^{[l_i]}(g_0^D))$ with multiplicities being equal
to dimensions of IRs $D^{[l_i]}(g_0^D)\, (D^{[l_i]}(G_i^0))$. In the case of
suitable (for given $H$ and $L(H)$) groups $G^0_i$ decompositions
(\ref{eq:SD}) have the simple spectra: $\sigma([l_i])=1$, and, then, pairs
$(G_i^0, g_0^D)$ (or $(G_i^0, G_0^D),\,G_0^D= \exp (g_0^D),\,G_i^0\otimes
G_0^D\subset G_0^{DM}$) say to act complementarily \cite{mq,al} on $L(H)$
and to form the Weyl-Howe dual pairs \cite{k93a} since pairs $(G^0_i=S_N,
G_0^D=U(n))$ of permutation and unitary groups were first considered
within quantum mechanics by H. Weyl \cite{w1}, and their explicit
mathematical characterization for pairs $(O(n), Sp(2m,R))$ of orthogonal
and symplectic groups was given by R. Howe \cite{how} (from hereon indices
$i, 0, D$ are omitted whenever it is of no importance). Note that implicitly
such Weyl-Howe dual pairs were used in different fields of many-body
physics (see, e.g., \cite{al,k93a} and references therein); without
dwelling on a review of these applications we mention some of known
examples: pairs ($SU(n),SU(m)$) in particle physics \cite{ok}, ($U(1),
su(1,1) \subset sp(2m,R))$($m \rightarrow \infty$) in superfluidity
theory \cite{per} and ($C_2, SU(1,1)$) in describing so-called squeezed
light \cite{s}.

The constructions above are generalized in a natural manner when extending
quadratic Hamiltonians $H_0$ by $G_i^I \,(\subseteq G_i^0$)-invariant 
polynomials $H_I(a_i, a_i^+,b_j, b_j^+)$ of higher degrees which describe
essentially nonlinear interactions \cite{k91,bg} (and, often, with enlarging 
Hilbert spaces $L(H)$). In general, such extensions lead to dual pairs where
dynamic algebras $g^D$ are infinite-dimensional graded Lie (super)algebras
$g^D=\sum_{r=-{\infty}}^{\infty} g_r,\,[g_r,g_s]\subset g_{r+s}$ enlarging
Lie algebras $g_0^D$ and embedded into enveloping algebras ${\cal U} (w(n;
m))$ of algebras $w(n;m)$ (from hereon we omit the subscript "$\pm$"
in $[,]_{\pm}$ whenever it is unnecessary) \cite{k88}. However
$G_i$-invariance of $H$ enables us to get generalized dual pairs $(G_i^I,
g^D= \hat g)$ where dynamic symmetry is described by finite-dimensional
non-linear (polynomial) Lie (super)algebras $\hat g= g_0^D + y_++ y_-$
extending Lie algebras $g_0^D$ and having an independent meaning. These
algebras $\hat g$ are introduced with
the help of $G_i^I$-invariant polynomial Jordan mappings which in the
simplest case, when $H_I(\dots)$ are homogeneous polynomials in $a_i, a_i^+,
b_j, b_j^+$, has the form $\tilde f$ \cite{k91,k88}
\begin{equation}
(a_i, a_i^+, b_j, b_j^+) \stackrel{\tilde f}\longmapsto (F_{\alpha},
Y_{\lambda}, Y^+_{\lambda}) \in \hat g, \quad  [g_0^D, y]\subseteq y,\quad
[y, y]\subset {\cal U}(g_0^D),\quad y= y_+ + y_-
\label{eq:pjm}
\end{equation}
where generators $Y_{\lambda}\in  y_-, Y^+_{\lambda}\in  y_+ $ are
simultaneously elementary vector $G_i$-invariants \cite{w1} and components
of two mutually conjugate $g_0^D$-irreducible tensor operators $Y, Y^+$.
In practice, Hamiltonians $H_0, H_I$ may be inhomogeneous polynomials
in $a_i, a_i^+, b_j, b_j^+$
and, besides, contain other $g_0^D$-covariant operators that leads to
modifications of Eq. (\ref{eq:pjm}) \cite{k93a,kk}. The first example
of using the mapping (\ref{eq:pjm}) in
physical problems was given (implicitly) in \cite{bg}  for extending the
unitary algebra $u(1)$ by its $C_n$-invariant symmetric tensors; later such
constructions were introduced explicitly in \cite {k88,k93a} for extending
algebras $u(m)$ by their $C_n$-invariant symmetric and $SU(n)$-invariant
skew-symmetric tensor operators (see Section 5) as well as for extending
the symplectic algebras $sp(2m,R)$ by $SO(n)$-invariant skew-symmetric
tensors.

Without dwelling on a complete analysis of the algebras $\hat g$ we outline
some of their features. As is seen from Eq. (\ref{eq:pjm}), algebras $\hat g$
resemble in their structure so-called $q$-deformed Lie algebras (widely used
for last time \cite{smp}) and have the coset structure (generalizing the
Cartan decomposition for real semisimple algebras \cite{bar}) that enables
us to construct IRs of $\hat g$ starting from $g_0^D$-modules. However,
unlike usual (linear) Lie algebras, exponentials $\exp (\hat g)$ generate
only pseudogroup structures rather than finite-dimensional Lie groups (cf.
\cite {km}) that impedes direct extensions of standard group-theoretical
techniques for solving physical tasks \cite {k97}. Nevertheless, using
generalizations \cite{k93a} 
\begin {equation}
(F_{\alpha},Y_{\lambda}, Y^+_{\lambda})\;\stackrel{\tilde f}\longmapsto\;
(F^0_{\alpha}=F_{\alpha},F'^+_{\alpha},F'_{\alpha})\in h, \quad
F'^+_{\alpha}= (F'_{\alpha})^+ = \sum_{\lambda} Y^+_{\lambda}
f_{\alpha,\lambda}(\{ F_{\beta}\}),\; [h, h]\subseteq h
\label{eq:HP}
\end{equation}
of the Holstein-Primakoff mappings \cite{jhp} (with $h$ being usual Lie
(super)algebras and "coefficients" $f_{\alpha,\beta} (\dots)$ determined
from sets of finite-difference equations), one can construct some
finite-dimensional Lie subgroups $\exp (h)\subset \exp (\hat g)$ which are
useful for physical applications \cite{k97}.

Let us now sketch some of physical aspects of formal constructions
above to elucidate the heuristic meaning of IDA. The key role belongs here
to the decomposition (\ref{eq:SD}) which describes "kinematic" premises of
arising CS in $L(H)$ due to the $G_i$ symmetry. Indeed, subspaces $L([l_i])$
consist of the "$g^D$-layers" $L([l_i];\nu)=Span\{|[l_i];\mu;\nu\rangle=
{\cal P}^{[l_i];D}_{\nu}(F^+_{\alpha}, Y_{\lambda}^+)|[l_i];\mu\rangle\}$
obtained by actions of polynomials in the $g^D$ positive weight shift
generators on basic vectors $|[l_i];\mu\rangle= {\cal P}^{[l_i];i}_{\mu}
|[l_i]\rangle$ of the IRs $D^{[l_i]}(G_i)$ which are simultaneously
specific (degenerated) "pseudovacuum" vectors with respect to $g^D:\,
Y_{\lambda}|[l_i];\mu\rangle= F_{\alpha}|[l_i];\mu\rangle= 0$. Thus,
$G_i$-invariance plays a "synergetic" role and yields "potential (kinematic)
forms" for CS which may be formed in $L(H)$ and are described by subspaces
$L([l_i])$ at the macroscopic level and by $g^D$-cluster variables
$F^+_{\alpha},Y^+_{\alpha}$ at the microscopic level. Note that, generally,
the decompositions (\ref{eq:SD}) contain the
"particular" ($G_i$-scalar) subspaces $L([0])$ "consisting" only of
$g^D$-clusters whereas other spaces $L([l_i])$ "contain" fixed (determined
by the "signatures" $[l_i]$) numbers of uncoupled or partially coupled
"primary particles". "Physical" realizations of these hidden CS are
implemented dynamically in their "pure" or "mixed" kinematic forms
determined by concrete $G_i$-invariant Hamiltonians $H_I$ (containing or
not $G_i$-covariant coupling parameters (fields) "mixing" different
$L([l_i])$) and initial states $|\psi(0)\rangle$. "Pure" realizations lead
to superselection rules for quantum numbers $l_i$ (cf. \cite{bar}) whereas
"mixed" ones imply possibilities of critical ("threshold") phenomena and
the spontaneous symmetry breaking (cf. \cite{izr}). And now we turn to
some recent examples of explicit IDA applications focusing our attention
only on key points.

\section{$G$-invariant bi-photons and the $SU(2)$-invariant treatment of
unpolarized light}
The first examples of applications of IDA to be examined deal with
quantum-optical parametric models with $m$ spatiotemporal and two
polarization ($\pm)$ light field modes whose Hamiltonians
\begin{equation}
H^2 =H_f +H_p^2,\quad H_f=\sum_{i=1}^{m} \omega_i \sum_{\alpha = +,-}
a_{\alpha i}^+a_{\alpha i},\;H_p^2= \sum_{i=1}^{m} \sum_{\alpha,\beta= +,-}
[g^{\alpha\beta}_{ij}a_{\alpha i}^+ a_{\beta j}^+ +g^{\alpha\beta *}_{ij}
a_{\alpha i} a_{\beta j}]
\label{eq:hsl}
\end{equation}
are quadratic in field operators and $c$-numbers $g^{\alpha\beta}_{ij}$
determine concrete parametric processes \cite{k91,k93a,k93p}. Their simplest
one-mode version ($m=1, \alpha=+(-)$) has the invariance group
$G_i^0=C_2=\{c_{k2}=\exp(i\pi k a^+ a),
k=0,1\}$ acting on the Fock space $L_F(1)\equiv L_F(1;0)=L(H^2)$ as
follows: $a^+\rightarrow c_{k2} a^+$. The dual pair is ($G_i=C_2, g^D=
su(1,1)= Span \{Y_0=a^+a/2+1/4, Y^+= a^{+2}/2, Y =a^2/2\}\sim Sp(2,R)$),
and the decomposition (\ref{eq:SD}) is trivial:\, $L_F(1)= L(0)+L(1/2)$
where the eigenvalue $l_0=\kappa/2$ of the operator $R_0=a^+a/2 -
[a^+a/2]$ ($[x]$ is the "entire part" of $x$), connected with the lowest
weights $k$ of the $su(1,1)$ IRs realized on $L_F(1):\kappa=2k-1/2$,
determines the number $N_{up}= \kappa$ of un-paired photons in $L(l_0)=
Span\{(Y^+)^{\mu} (a+)^{\kappa} |0\rangle\}$. The "particular" space $L(0)$
consists of bi-photons $Y^+$ and contains states $|\beta\rangle=
\exp (\beta Y^+-\beta^* Y)|0\rangle$ of the so-called "squeezed vacuum"
light \cite{s}. However, more interesting examples of CS in quantum optics
due to symmetry have been found recently by using a specific polarization
invariance of light fields.

Indeed, the free field Hamiltonian $H_f$ in (\ref{eq:hsl}) is invariant
with respect to the group $G_i^0=\prod_{i=1}^m U^i(2)\subset Sp(4m,R)$ where
$U^i(2)=\{\exp (i\gamma N_i+ i\eta_0 P_0(i) + \eta_ 1 P_+(i) -\eta_1^*
P_-(i))\}, N_i=\sum_{\alpha =+,-} N_{\alpha i} (N_{\alpha i}=a_{\alpha i} ^+
a_{\alpha i}$) is the photon number operator of the $i$-th spatiotemporal
mode and $P_0(i)=[N_{+i}-N_{-i}]/2,\,P_\pm (i)= a_{\pm i}^+a_{\mp i}$ are
generators of the $SU(2)^i_p \subset U(2)^i$ subgroups defining the
polarization $P(i)$-quasispins (related to the polarization Stokes vector
operators of single spatiotemporal modes) \cite{k93p}.
The group $G_i^0$ contains the $SU(2)_p$ subgroup generated by the total
$P$-quasispin operators $P_{\alpha} =\sum _{i=1}^{m}P_{\alpha}(i)$ and
enabled us to reveal hidden CS and to examine new collective phenomena
connected with "polarization clusterizations" of light field modes
\cite{k91,k93p}.

Really, the $SU(2)_p$ group acts on $L_F(2m)\equiv L_F(2m;0)$ complementarily
to the $so^*(2m)$ algebra generated by operators $E_{ij} \equiv\sum_{\alpha =
\pm} a^+_{\alpha i}a_{\alpha j}\in u(m)$ and $SU(2)_p$-invariants
$X^{+}_{ij}=a^+_{+ i}a^+_{- j}-a_{- i}^+ a^+_{+ j}:\,[P_{\alpha},
X^{+}_{ij}]=0, \alpha=0,+,-, X_{ij}=(X^{+}_{ij})^+$. The decomposition
(\ref{eq:SD}) of $L(H)= L_F(2m)$ with respect to the dual pair ($G_i^p=
SU(2)_p, g^D= so^*(2m)$) contains the infinite number of the $SU(2)_p\otimes
SO^*(2m)$-invariant subspaces $L(l_0=p)=L(p)=Span\{|p;\mu;\nu\rangle\}$
labeled by values $p$ of the total $P$-quasispin which also determine the
Casimir operator values of the $so^*(2m)$ IRs realized on $L_F(2m)$ and
are measured in experiments with "polarization noises" \cite{k93p,k91}.
Basic vectors $|p;\mu;\nu\equiv [n_i,p_j]\rangle$, specified by the $P_0$
eigenvalue $\mu$ (helicity), photon numbers $n_i$ and "intermediate" cluster
quasispins $p_j$, have, in general, the form $|p;\mu;\nu\rangle =
{\cal P}^{p;so^*(2m)}_{\nu}(X^{+}_{ij})|p;\mu\rangle$ where the $so^*(2m)$
"pseudovacuum" vectors $|p;\mu\rangle={\cal P}^{p;su(2)_p}_{\mu}(a_{\pm i}^+,
Y_{ij}^{+})|0\rangle$ are given by polynomials in $a_{\pm i}^+$ and
$P_0$-invariant operators $Y^{+}_{ij}=(a^+_{+ i}a^+_{- j}+a_{- i}^+
a^+_{+ j})/2:\,[P_0,Y^{+}_{ij}]=0$ which are direct analogs of Bogolubov's
pairs in superfluidity. The operators $Y^{+}_{ij},Y_{ij}=(Y^+_{ij})^+$
extend the algebra $so^*(2m)$ to the algebra $u(m,m)$ acting on $L_F(2m)$
complementarily to the polarization subalgebra $u(1)_p=Span\{P_0\}$. From
the physical point of view quantities $X^{+}_{ij},Y^{+}_{ij}$ may be
interpreted, respectively, as creation
operators of $P$-scalar and $P_0$-scalar bi-photon kinematic clusters
determining, in fact, two classes of unpolarized light (UL) associated,
respectively, with the "particular" subspaces $L(0)\equiv L(p=0)$ and
$L'(0)\equiv L'(\mu=0)=Span\{|p;\mu=0;\nu\rangle\}$ \cite{k93p}.

Indeed, in \cite{k93p} we proved that quantum states $|\rangle
\in L(0),\, L'(0)$ satisfy the familiar definition of UL: ${\cal P}\propto
[<P_0>^2+<P_1>^2+<P_2>^2]^{1/2}=0$ (${\cal P}$ is the light polarization
degree, $P_{\pm}=(P_1 \pm iP_2)$, the symbol $<\dots>$ denotes  both
statistical and quantum averages) and, besides,  extra
(polarization "classicality" and "squeezing") conditions:
\begin{equation}
a) <|P^s_{i=1,2,0}|>=0 \quad\forall s\geq2, \;| \rangle \in L(0);
\qquad b) <|P^s_0|>=0 \quad\forall s\geq2,\; | \rangle \in L'(0)
\label{eq:qUL}
\end{equation}
States $|\rangle\in L(0),|\rangle\in L'(0) (P$-and $P_0$-scalar
light in terminology \cite{k93p}) are natural (and "particular" due to Eqs.
(\ref{eq:qUL})) representatives of two (introduced in \cite{llp} and named
as $P$-and $P_0$-invariant light in \cite{k96p}) kinds of UL which obey
general invariance conditions used in \cite{pch,llp} (in different forms)
for more strong (in comparison with the above familiar) definitions  of UL
retaining some features of the natural (thermal) UL. Namely, states of
$P$-invariant light satisfy the conditions
\begin{equation}
a)\mbox{Tr}[S\rho S^{\dagger} A(\{P_{\alpha}\})]=Tr[\rho A(\{P_{\alpha}\})]
\equiv <A(\{P_{\alpha}\})> \qquad \Longleftrightarrow \qquad b) S\rho
S^{\dagger}= \rho
\label{eq:ULd}
\end{equation}
for arbitrary $P_{\alpha}$-dependent observables $A(\{P_{\alpha}\})$ or
field density operators $\rho$ (and appropriate quasiprobability functions)
with any $S=\exp(ib_0 P_0 +b_1 P_+-b_1^* P_-)\in SU(2)_p$ while states of
$P_0$-invariant light obey Eqs. (\ref{eq:ULd}) with $S=\exp(ib_0 P_0),
\exp(i\pi P_2)\in SU(2)_p$. Emphasize, however, that $P_0$-and $P$-scalar
types of UL are due to strong phase correlations between photons
unlike familiar states of UL generated by randomizing mechanisms. Note also
that, in fact, the usual defiinition of arbitrary UL states (${\cal P} =0$)
can be given
in the form (\ref{eq:ULd}a) with any $S\in SU(2)_p$ if taking in it {\bf
only linear} functions $A(\{P_{\alpha}\})$ \cite{k96p}. All these
observations lead to a new treatment of (quantum
and classical) UL states based on their $SU(2)_p$ invariance properties and
to a natural division UL into two classes: 1) the weak UL having a
characteristic property (\ref{eq:ULd}a) with any $S\in SU(2)_p$ only for
first moments $<P_{\alpha}>$ (measured in standard polarization experiments)
and 2) the strong UL possessing invariance properties (\ref{eq:ULd}) for
higher moments and including $P_0$-and $P$-invariant light.

So, taking into account only the $SU(2)_p$ invariance of $H_f$ we have found
in $L_F(2m)$ hidden kinematic CS ("polarization domains") described by
subspaces $L(p)$ and $L'(\mu)$ which, according to general remarks of
Section 2, can be realized "physically" with the help of $G_i^p$-invariant
interaction Hamiltonians $H_I$ of two kinds: \,1) $H_I=H_I^{X,Y}$ depending
only on bi-photon variables $Y_{ij},Y^+_{ij}; X_{ij}, X^+_{ij}$ and
$G_i^p$-scalar coupling constants;\,2) $H_I=H_I^{cov}$ containing "free"
photon operators $a^+_{\pm i}$ and $G_i^p$-covariant coupling parameters
describing (phenomenologically) the chiral $SU(2)$-symmetry of the matter
(that, perhaps, is realized in some of
biophysical models) \cite{k93p}. The simplest examples of
$H_I^{X,Y}$ are obtained from Eqs. (\ref{eq:hsl}) by imposing conditions:
$g^{+-}_{ij}=\mp g^{-+}_{ij}=\tilde g_{ij}$ and $g^{\alpha\beta}_{ij}=0$
otherwise in $H_f$; actually, their $(X_{ij}, X^+_{ij})$-independent versions
were used for producing $P_0$-scalar light (as states $\exp (\beta Y^+_{11}-
\beta^* Y_{11})|0\rangle$ of the "two-mode squeezed vacuum" \cite{s}) while
the problem of an experimental production of $P$-scalar light is not yet
solved \cite{k93p}.
\section{Coherent clusters in nonlinear models of quantum optics}
The examples of applications of IDA using generalized Weyl-Howe dual pairs
$(G_{i}, \hat g)$ are yielded by generalizations of models (\ref{eq:hsl})
describing multiphoton scattering processes and quantum matter-radiation
interactions \cite{k93a}; their simplest versions are
given by Hamiltonians
\begin{equation}
H_{mp} = \sum _{i=1}^m \omega_i a_i^+ a_i  + \omega_0 a_0^+ a_0 +
\sum_{1\leq i_1\leq i_2\dots\leq i_n\leq m} [g_{i_1\dots} (a^+_{i_1}\dots
 a^+_{i_n}) a_0 + g_{i_1\dots}^*(a_{i_1}\dots a_{i_n}) a^+_0],\quad n\geq 2
\label{eq:cf}
\end{equation}
where polarization labels are omitted in subscripts $"i"$ and non-quadratic
parts of $H_{mp}$ describe, in particular, higher harmonics generation
($H_{mp}=H_{hg}$ when $m=1$) and frequency conversions ($H_{mp} =H_{fc}$
when $m=n$ and only $g_{12\dots n}\neq0$) whereas models of matter-radiation
interactions are obtained via replacing in Eq. (\ref{eq:cf}) the "pump"
mode $a^+_0$ by "atomic" operators \cite{k93a,kk}.

The general Hamiltonians (\ref{eq:cf}) have the invariance groups $G_i^{mp} =
C_n=\{\exp(i2\pi k a^+ a/n), k=0,1,\dots,n-1\}\subset \prod_j U^j(1)=
\exp(i\lambda_j  a^+_j a_j)=G_i^0$ whereas their specifications may have
extra factors $\exp(i\beta_j R_j)$ related to dynamic constants (integrals
of motion) $R_j\in Span\{N_i=a^+_i a_i\}$ describing additional interaction
symmetries; for instance, models $H_{hg}$ have dynamic constants $R_1=(N_1+
n N_0)/(1+n)$. Groups $G_i^{mp}$ form on the Fock spaces $L_F(n+1)$
generalized dual pairs $(G_i^{mp}, g_{mp}^D=\hat g^Y(n+1))$ together with
polynomial Lie algebras $\hat g^Y(n+1)= u(n+1) + y(n;1)$ obtained via the
mapping (\ref{eq:pjm}) as extensions of the Lie algebras $u(n +1)=
Span\{E_{ij}=a^+_i a_j\}$ by coset spaces $y(n;1)=Span\{Y^+_{i_1...i_n;0}=
a^+_{i_1}...a^+_{i_n} a_0, Y_{i_1...i_n;0}=(Y^+_{i_1...i_n;0})^+\}$; herewith
commutators $[Y_{i_1...i_n;0},Y^+_{j_1...j_n;0}]$ are polynomials in $E_{ij}$
\cite{k88}. Such an introduction of $G_i^{mp}$-invariant collective variables
$E_{ij}, Y^+_{\dots;0},Y_{\dots;0}$ enables us to rewrite Hamiltonians
$H_{mp}$ in the linear form (\ref{eq:H0}) with respect to $E_{ij},
Y_{\dots;0}, Y^+_{\dots;0}$ and to apply the $\hat g^Y(n+1)$ formalism for
revealing hidden CS and examining collective dynamic peculiarities in models
(\ref{eq:cf}) which slip off within standard studies \cite{k91,k96a}.

In order to elucidate basic ideas of such applications we restrict our
analysis by models with Hamiltonians $H_{hg}$ when $\hat g^Y(n+1)$ are
reduced to the polynomial Lie algebras $su_{pd}(2)= Span\{Y_{0}=
(N_1-N_0)/(1+n), Y_+=(a^+_1)^n a_0, Y_-=(Y_+)^+\}$ with commutation
relations
\begin{equation}
[Y_0, Y_{\pm}]=\pm Y_{\pm},\; [Y_-,Y_+]=\Phi (Y_0;R_1)\equiv\Psi (Y_0+1;R_1)
-\Psi (Y_0;R_1), \quad [Y_{\alpha}, R_1]=0
\label{eq:crV}
\end{equation}
resembling those for $su(2)$ but with polynomial structure functions
$\Psi (Y_0;R_1)=(R_1-Y_0+1)(nY_0+R_1)^{(n)} (A^{(n)}\equiv A(A-1)\dots
(A-n+1)$) \cite{kk}; note that $R_1$-dependence of $\Phi (Y_0;R_1)$, in fact,
"intertwines" $G_i^{hg}=C_n\otimes\exp(i\beta R_1)$ and $g^D=su_{pd}(2)$ in
an algebraic object resembling the semidirect product of groups (cf.
\cite{ed,bar}). Then Hamiltonians $H_{hg}$ are expressed by linear functions
\begin{equation}
H_{hg}= aY_0+ b Y_++b^*Y_-+ c R_1, \quad a=
n\omega_1-\omega_0, \, b= g_{1\dots 1},\,c=(\omega_1+\omega_0)
\label{eq:hpa}
\end{equation}
in the generators $Y_{\alpha}$ and dynamic constant $R_1$, and the
decomposition (\ref{eq:SD}) of $L(H)=L_F(n+1)$ with respect to
($G_i^{hg}, su_{pd}(2)$) contains the infinite number of the
$su_{pd}(2)$-irreducible $s$-dimensional subspaces $L([l_i])=
Span\{(Y_+)^{\eta}|[l_0]\rangle,\,|[l_0]\rangle=(a^+_1)^{\kappa}(a^+_0)^s
|0\rangle, \kappa=0,\dots,n-1, s\geq 0 \}$ labeled by eigenvalues
$l_0=(\kappa -s)/(1+n),l_1= (\kappa +ns)/(1+n)$ of $R_i$ where
$R_0$ is determined from the identity: $\Psi (R_0;R_1)\equiv \Psi (Y_0;R_1)-
Y_+ Y_-$ defining the $su_{pd}(2)$ Casimir operator \cite{k96a}.

This "$su_{pd}(2)$-cluster" formulation of models entails a dimension
reduction of physical tasks and an explicit "geometrization" of model
dynamics manifesting already at the classical level of examination.
So, e.g., the decomposition (\ref{eq:SD}) implies the representaion
of model phase spaces $C^{n+1}$ as fiber bundles: $\,C^{n+1}=
\cup_{[l_i]}{\cal A}([l_i])$ where $su_{pd}(2)$-invariant dynamic manifolds
${\cal A}([l_i])$ are Abelian varieties corresponding to spaces $L([l_i])$
and given (using the mean-field approximation) in dynamic variables
$\bar Y_{\alpha}=<Y_{\alpha}>$ as follows: ${\cal A}([l_i]) =
\{\bar Y_{\alpha}: 2\bar Y_+\bar Y_-= \Psi (\bar Y_0; l_1) +\Psi (\bar Y_0+1;
l_1)\}$. Then, states belonging to a fixed manifold ${\cal A}([l_i])$ (or
${\cal A}([\bar R_i])$ in the general case) will evolve in it under
action of Hamiltonian flows with Hamiltonian functions ${\cal H}=a \bar Y_0+
b\bar Y_++b^*\bar Y_-+ c \bar R_1$. Herewith (approximate) dynamic
trajectories are determined as intersections of manifolds
${\cal A}([\bar R_i])$ with energy planes ${\cal H}=E$ that enables us
to determine some peculiarities of model dynamics \cite{k96a}.

These considerations become more transparent if
using "quasi-spin" reformulations of the models (\ref{eq:hpa}) in terms of
the $su(2)$ generators $V_{\alpha}$ connected with $Y_{\alpha}$ via the
mapping (\ref{eq:HP}): $V_0= Y_0-R_0- J,\, V_+=Y_+[\varphi (V_0)]^{1/2},
\varphi (V_0)=(J+V_0+1)(J- V_0)/\Psi (Y_0+1;R_1),\,Y_-=(Y_+)^+ $ ($J$ is
the $su(2)$ highest weight operator with eigenvalues $j=s/2$)\cite{k93a}.
Then the Hamiltonians (\ref{eq:hpa}) are represented by nonlinear functions
\begin{equation}
H= aV_0+ b V_+ [\varphi (V_0)]^{-1/2}+b^*[\varphi (V_0)]^{-1/2}V_-+
cR_1+ a(R_0+J)
\label{eq:hqs}
\end{equation}
in the "$su(2)$-cluster" variables $V_{\alpha}$, and fiber bundle
representations of phase spaces $C^{n+1}$ contain $SU(2)$-invariant "Bloch
spheres" $S^2_{j}: \bar V_0^2 +\bar V_1^2 +\bar V_2^2 =j^2$ instead of
$su_{pd}(2)$-invariant manifolds ${\cal A}([l_i])$  while energy
planes are replaced by nonlinear energy surfaces $<H>=E$. Furthermore,
these "quasi-spin" reformulations enable
us to get  new (in comparison with obtained earlier) $su(2)$-cluster
quasiclassical solutions of spectral and evolution tasks
using techniques of the $SU(2)$ coherent states $|\phi_0; \alpha>=
S_V(\alpha)|\phi_0>\in L(H),\, S_V(\alpha)=\exp(\alpha V_+ -\alpha^* V_-)$
which can be of "spin-like" type (when $|\phi_0>\in L([l_i])$) or $su(2)$-
reducibile (when $|\phi_0>\in L(H)$) \cite{k96a}.

For example, energy eigenstates $|E([l_i];v)\rangle$ and spectra
$\{E([l_i];v)\}$ can be approximated by means of standard variational schemes
with using $SU(2)$ coherent states $ S_V(\xi) {\cal N}([l_i];v) V_+^v
|[l_i]\rangle=|[l_i];v;\xi\rangle$ as trial functions.
Namely, we find approximate
eigenstates $|E^{qc}([l_i];v)\rangle=|[l_i];v;\xi\rangle$ and eigenenergies
$E^{qc}([l_i];v)= \langle[l_i];v;\xi|H|[l_i];v;\xi\rangle$ where values
of the parameter $\xi= r\exp (-i\theta)$ are determined by the stationarity
conditions
\begin{equation}
\frac{\partial {\cal H}([l_i];v;\xi)}{\partial \theta}=0,\qquad 
\frac{\partial {\cal H}([l_i];v;\xi)}{\partial r}=0, \quad
{\cal H}([l_i];v;\xi)=\langle[l_i];v;\xi|H|[l_i];v;\xi\rangle
\label{eq:Hev}
\end{equation}
for the energy functional ${\cal H}([l_i];v;\xi)$.
In fact, in such a way we get $\exp (-i\theta)=b/|b|$ and a whole series of
competitive potential solutions for values $r$; their final selection
may be made with the help of a "quality criterion" using the "energy error"
functionals introduced in \cite{k93a}. Similarly, an appropriate
quasiclassical dynamics is described by the classical Hamiltonian
equations \cite{k96a}
\begin{equation}
 \dot q =\frac{\partial {\cal H}}{\partial p}, \qquad \dot p =
-\frac{\partial {\cal H}}{\partial q}, \quad {\cal H}=
\langle \phi_0; z(t)|H|[\phi_0;z(t)\rangle ,\quad q= \theta,\quad p\equiv
\langle \phi_0; z(t)|Y_0|\phi_0;z(t)\rangle
\label{eq:Hfl1}
\end{equation}
for "motion" of the canonical parameters $p,q$ of the $SU(2)$ coherent
states $|\phi_0;z(t)\rangle=S_V(z(t))|\phi_0\rangle$ ($z=- r
\exp (i\theta)$) as trial functions
in the time-dependent Hartree-Fock variational scheme.
Note that solutions of Eqs.(\ref{eq:Hev})-(\ref{eq:Hfl1}) smoothly
approximate exact ones and catch explicitly quantum cooperative features
of models at the quasiclassical levels \cite{k96a}.

\section{Generalized dual pairs in the theory of composite fields}

Another area of a "natural" appearence of generalized dual pairs $(G_i,
g^{DS}=\hat g)$ is the algebraic analysis \cite{k91,k97} of composite
fields with internal (gauge) symmetries \cite{bar} which generalizes
basic ideas of the paraquantization \cite{ohk,bke} and implements in a
sense the method of fusion by L. de Broglie \cite{brl}. Actually, the
simplest example of such an analysis (but without introducing dual pairs
and non-linear Lie algebras $\hat g$)  was given in \cite{bg} by means
of using $n$-boson one-mode versions
\begin{equation}
H^n =  \omega_1 a_1^+ a_1 + g Y^+_{1\dots} + g^* Y_{1\dots},
\quad Y^+_{1\dots}=a_{1}^+\dots a_{1}^+ = (a^+_1)^n
\label{eq:hcm}
\end{equation}
of Hamiltonians (\ref{eq:hsl}) to describe resonance states in particle
physics; later it was generalized on multimode cases to study multiphoton
processes in quantum optics (see \cite{k91} and references therein).

Specifically, in \cite{bg} it was shown that operators $Y^+\equiv
Y^+_{1\dots}$ describe  $n$-particle kinematic clusters which display
unusual (para)statistics and correspond to generalized asymptotically
free fields realized on the Fock space $L_F(1)$. In fact, the operators
$Y^+, Y=(Y^+)^+, Y_0=a_1^+a_1/n\equiv E_{11}/n$ satisfy \cite{k91}
(non-canonical) commutataion relations (\ref{eq:crV}) of the $su_{pd}(1,1)$
algebra with the structure polynomial $\Psi (Y_0) =(E_{11})^{(n)}$ and,
besides, extra multi-linear relations:
$ad^{n+1}_{Y}Y^+=ad^n_{Y}(ad_{Y}Y^+)=0,\, ad_{Y}Y^+\equiv [Y, Y ^+]$,
generalizing (for $n\geq 3$) trilinear parastatistical Green's relations
\cite{ohk,bke}. Thus, we get an action of the generalized dual pair $(G_i=
C_n =\{\exp(i2\pi k a^+_1a_1)/n\},\hat g=su_{pd}(1,1))$ on the space
$L_F(1)$.  The appropriate decomposition (\ref{eq:SD}) contains the
subspaces $L([l_0=\kappa/n])= Span\{(Y^+)^{\eta}|[l_0]\rangle,\,
|[l_0]\rangle=(a+)^{\kappa}|0\rangle\}, \kappa=0, 1,\dots,n-1,$ describing
coherent mixtures of constant numbers $\kappa$ of uncoupled bosons $a_1^+$
and of varying in time numbers $N_Y$ of $Y$-clusters. However,
operators $N_Y$ have not standard (for (papra)fields) bilinear in $Y, Y^+$
forms \cite{ohk} but they can be expressed (due to the evident identity
$\Psi (Y_0) = Y^+Y$ on $L_F(1)$) as nonlinear functions in the bilineals
$Y^+Y, YY^+$  \cite{k91}:
$N_Y =(E_{11}- nR_0)/n= [{a^+_1a_1}/n]=[{E_{11}}/n],\, E_{11}=
\varphi (Y^+Y)\equiv n\Psi^{-1} (Y_0)$
as it is the case for algebras $A(K)$ describing non-standard statistics 
\cite{bke}. Therefore, at best the
quantities $Y^+,Y$ can be set in correspondence only
to parafield (when $n=2$) quanta \cite{ohk,bke} rather than to certain
asymptotically free particles \cite{k91}. Nevertheless, one can construct
from them operators $W^+=W^+(\{Y_i\}), W=(W^+)^+$ obeying canonical
commutation relations $[W, W^+]=1$, having the standard number operators
$N_W=W^+W(=N_Y)$ and corresponding to quanta of asymptotically free
multi-boson fields (which can be realized in subspaces $L([0])$ in "pure
 forms"). Actually, two equivalent forms 
\cite{k91,bg,k93a}:
\begin{equation}
W^+ = Y^+\sum_{r\geq 0} c_r (Y^+)^r(Y)^r= \,Y^+ [(Y_0-R_0+1)
/(E_{11}+n)^{(n)})]^{1/2}, \quad W=(W^+)^+
\label{eq:pco1}                                             
\end{equation}
were found for such $W^+,W$ where the second one is, a specification of the
mapping (\ref{eq:HP}).

The analysis above has been generalized \cite{k88} by means of: 1) using
$"m"$-mode extensions of models (\ref{eq:hcm}) with $C_n$-invariant
interaction Hamiltonians $ \sum_{1\leq i_1\dots\leq m}[g_{i_1\dots}
Y^+_{i_1\dots} + g^*_{i_1\dots} Y_{i_1\dots}], \,Y^+_{i_1\dots}=
a_{i_1}^+\dots a_{i_n}^+$; 2) considering their analogs with non-Abelian
groups $G_i=SU(n)$ (whose Hamiltonians are obtained by the substitutions:
$a_i^+ a_i\rightarrow ({\bf a}^+_i \cdot{\bf a}_i)\equiv\sum_{j=1}^n
a_{ji}^+ a_{ji},\; Y^+_{i_1\dots}\rightarrow X^+_{i_1\dots}\equiv
\sum_{(j_k)} \epsilon_{j_1 \dots j_n}
a_{j_1 i_1}^+\dots  a_{j_n i_n}^+$ ($\epsilon_{j_1 \dots j_n}$ is the
totally antisymmetric tensor); 3) involving both boson and fermion variables.
These procedures yield
a variety of generalized dual pairs; for instance, when using
two first ones we get dual pairs ($C_n, {osc}^{Y}(m;(n))$) and ($SU(n),
{osc}^{X}(m;1^n)$) where ${osc}^{Y}(m;(n))$ and ${osc}^{X}(m;1^n)$ are
extensions of the unitary algebras $u(m)=Span\{ E_{ij},\,
E_{ij}=a_i^+ a_j \,\mbox{and}\, E_{ij}=({\bf a}^+_i\cdot{\bf a}_i)\}$ by
their symmetric ($Y^+_{\dots}, Y_{\dots}$) and skew-symmetric ($X^+_{\dots},
X_{\dots}$) tensor operators \cite {k88,k97}. The operators
$X^+_{\dots},X_{\dots}$ and $Y^+_{\dots},Y_{\dots}$ satisfy non-canonical
commutation relations whose right sides depend on $E_{ij}$ (and on the
$SU(n)$ Casimir operators for $osc^{X}(m;1^n)$) and obey (due to the
invariant theory \cite{w1,how}) certain extra "bootstrap" relations
("syzygies") of the type: $Y_{1\dots1} Y_{2\dots2}=Y_{21\dots1}
Y_{12\dots2}$ \cite{k88,k91} which are similar to those occuring in quantum
field theories with constraints \cite{bar,ohk} and in non-standard
quantization schemes dicussed in \cite{bke}. All this entails unusual
statistical and other features  of $G_i$-invariant clusters associated
with $X^+_{\dots}, Y^+_{\dots}$ and complicates extensions of the one-mode
analysis above \cite{k88}. Specifically, the task of
obtaining $m$-mode generalizations 
\begin{equation}
W^+_a = \sum_{i_1\dots i_n} (Y^+_{i_1\dots i_n}/X^+_{i_1\dots i_n} )
f^a_{i_1\dots i_n} (\{ E_{ij}\}),\qquad [W_a, W^+_b]=\delta_{a b},\; W_a=
(W^+_a)^+
\label{eq:pco2}
\end{equation}
of the mapping (\ref{eq:pco1}) is, in general, fairly difficult owing to
"syzygies" between $Y/X$-clusters (and resembles the "reducibility problem"
for algebras $A(K)$ \cite{bke}).

When determining explicit expressions for $f_{\dots}(\dots)$ in Eqs.
(\ref{eq:pco2}) (and in their generalizations, e.g., for constructing $W^+_a
\in A(K)$) we get an effective tool for analyzing composite field
models with internal $G_i$-symmetries at the
algebraic and quasi-particle levels (including a new insight into some "old
problems", such as, e.g., the quark confinement \cite{ed,ok}).
Furthermore, examining the limit "$m \rightarrow\infty$" and
involving spatiotemporal variables and symmetries into consideration, one 
can also construct in terms of "quanta" $W_a$ appropriate  "physical" 
(asymptotically free) composite fields \cite{k91} and, then, develop for them
standard theories including non-linear (due to Hamiltonian forms) evolution 
equations and their soliton/instanton solutions \cite{izr}; herewith
discrete quantum numbers $l_i$ labeling subspaces $L([l_i])$ in
(\ref{eq:SD}) may display themselves as specific topological charges.
In particular, in such a way, using suitable analogs of Eq. (\ref{eq:pco1}) 
for $P/P_0$-scalar biphotons \cite{k93p}, we answer in the affirmative
within quantum optics the problem of existence of UL waves put by A. Fresnel
in the beginning of XIX century and having the negative solution within
the framework of classical electrodynamics due to the vector
nature of the Maxwell equations \cite{li,bl1}.
\section{Conclusion}
So, we formulated mathematical grounds of IDA and showed its physical meaning
"in action". In conclusion we briefly discuss some ways of applying and
developing results obtained.

The general constructions of Sections 2,5 may be applied for the systematic
search of hidden CS within different areas of quantum many-body physics
by using known dual pairs \cite{al,k93a} and for developing field theories
with "hidden quantum variables" and unusual statistics \cite{ohk,k91,bke}
(including the problem of consistency of the Poincare symmetry with dynamic
ones \cite {izr,om,k91}). On other hand, they are useful in solving
appropriate "inverse problems" \cite{ksh}: to display hidden symmetries
$G_i$ and "pre-particles" from analyzing spectroscopic data for
complex systems associated with IRs of certain dynamic algebras $g^D$ (that
is of very importance when interaction Hamiltonians are determined
phenomenologically). For this aim it is worth-while to enlarge lists of
dual pairs used by involving new classes of groups $G_i$ and q-deformed
oscillators  into consideration \cite{k97}.

More concrete results of Sections 3,4, firstly, can be used as general
patterns of applying IDA in $G_i$-invariant many-body models and,
secondly, open new lines of investigations in quantum optics. So, e.g.,
the above $SU(2)_p$-invariant treatment of UL stimulates experiments on
producing new states of quantum UL (especially, of $P$-scalar light),
studies of interactions of these states with material media \cite{k91}
and their applications in communication theory, spectroscopy of anisotropic
media and biophysics \cite{k93p}. At the same time "quasi-spin" formulations
and $su(2)$-cluster quasiclassical approximations in models (\ref{eq:cf})
outline (related to  geometric quantization schemes \cite{km}) ways of
"geometrization" of dynamics in models of strongly interacting subsystems
and, simultaneously, can be used to reveal new collective phenomena in
such models, including topological features of Hamiltonian flows determined
by Eqs. (\ref{eq:Hfl1}) at the different quasiclassical levels \cite{k96a}.

The author thanks G.S. Pogosyan for his interest and attention to the work
and V.P. Bykov, R.N. Faustov and S.M. Chumakov for  useful discussions.
A support of the work from the Russian State Research and Technology
Program "Fundamental Spectroscopy" and the Russian Basic Research
Foundation (grant No96-02 18746a) is acknowledged.

\end{document}